\documentclass[conference]{IEEEtran}

\usepackage{amsmath}
\usepackage{listings} 
\usepackage{url}
\usepackage{graphicx}

\begin{document}

\title{Alleviating Merge Conflicts with Fine-grained Visual Awareness}

\author{\IEEEauthorblockN{Stanislav Levin}
\IEEEauthorblockA{The Blavatnik School of Computer Science\\
 Tel Aviv University\\
Tel-Aviv, Israel\\
stanisl@post.tau.ac.il}
\and
\IEEEauthorblockN{Amiram Yehudai}
\IEEEauthorblockA{The Blavatnik School of Computer Science\\
 Tel Aviv University\\
Tel-Aviv, Israel\\
amiramy@tau.ac.il}
}

\maketitle

\begin{abstract}

Merge conflicts created by software team members working on the same code can be costly to resolve, and adversely affect productivity. In this work, we suggest the approach of fine-grained merge conflict awareness, where software team members are notified of potential merge conflicts via graphical decoration of the relevant semantic elements, in near real-time. The novelty of this approach is that it allows software developers to pinpoint the element in conflict, such as a method's body, parameter, return value, and so on, promoting communication about conflicting changes soon after they take place and on a semantic level. We have also conducted a preliminary qualitative evaluation of our approach, the results of which we report in this paper.

\end{abstract}


\section{Introduction}

Modern software projects involve multiple developers collaboratively working on the same codebase. In fact, parallel development has become the norm rather than the exception \cite{SupportingIndirectConflicts}. The task of sharing a codebase repository is usually carried out by a Software Configuration Management system (SCM) \cite{SVN, ClearCase, GIT, Mercurial}. The SCM system maintains all files that comprise the software project, and serves as the only version controlling mechanism through which developers share code \cite{EnvironmentForSyncDev}. The SCM tools employ a common checkin / checkout model according to which a change will become visible to others, only after the developer who made it checks in his code to the shared repository. 

A direct implication of this model is that using an SCM system alone, without additional tools, conflicting code changes will only be discovered post factum, when a developer tries to checkin the already conflicting code. Once aware of the conflict the developer is forced to resolve it by means of merging his version with the repository's one. Such manual merges are considered both time consuming and error prone \cite{ConcurrencyControlInGroupwareSystems, ProactiveConflictDetection}. Since the conflict involves changes made by multiple team members, in order to make the correct decision a comprehensive understanding of the overall changes must be obtained. The process of obtaining the information pertaining to each change may be done in various manners. For instance, one can query fellow developers about the changes they made; or inspect the history log, often provided by the SCM system; some methods even provide an inherent support for dealing with conflicts, such as the multi-versioning technique described in \cite{MultiVersionConflictResolution}. Regardless of the method chosen, one thing remains painfully certain - a mishandled merge can lead to a variety of negative results, ranging from ending up with code that generates compilation errors in the shared repository, to an evident faulty program behavior, or even worse, an inevident faulty program behavior, that will only be discovered later on, well after the merge activity and its delicate context are a thing of the past. It should be noted that the term \textit{merge} is used both for the task of merging conflicting, and non conflicting, code changes. This observation plays an important role, since although both cases deal with merging several versions (or changes) into one, the challenges involved in each of them are of different nature. Merging conflicting changes is a much more intricate procedure.

Distributed SCM tools (e.g., Git \cite{GIT}, Mercurial \cite{Mercurial}) have suggested improved methods to efficiently and automatically merge non conflicting changes \cite{MovingToDSCM} and thus alleviate the task of merging. However, \cite{ProactiveConflictDetection} concluded that even with these modern SCM systems, merge conflicts are frequent, persistent, and appear not only as overlapping textual edits but also as subsequent build and test failures. 

Merging conflicting changes presents an even harder problem, one that extends beyond technical difficulties, and is unresolvable by automatic means since there is no right or wrong, it just so happened that several changes had taken place simultaneously and affected the same element. Each change is syntactically and semantically valid, but only one can be included in the final code version. Software development practices such as Continuous Integration \cite{duvall2007continuous, fowler2006continuous} can facilitate early detection of compilation errors and faulty program behavior (if it is covered by automatic tests), by automatically performing periodic build cycles that include both compilation and testing stages. However, it is often not early enough, since a conflict that has already made its way into the codebase severely disrupts the normal development workflow. If the code in the shared repository is misbehaving, anyone who updates his private copy will get the faulty behavior as well, possibly without even being aware of it.

\label{reactiveAndProactive} Most methods suggested in recent years deal with conflicts re-actively, i.e., once they detect a potential conflict they \textit{react} by propagating notifications in an attempt to make team members aware of the conflict, yet they do not necessarily make it their goal to prevent it in the first place. Synchronized Software Development (SSD) \cite{2015arXiv150406742L} was a pioneer approach that was designed to behave pro-actively, and try to \textit{prevent} conflicts before they can take place, by implementing a fine-grained semantic locking \cite{slevin_thesis}, where concurrent editing of the same semantic element was not allowed.

\section{Related work}
A number tools have been suggested to facilitate merge conflicts introduced by the checkin / checkout model, and they can be classified into two dominant categories, according to whether they insert code from external sources into one's private copy of the code, or not. We call the former an \textit{intrusive} strategy, and the latter a \textit{non-intrusive} strategy.

\subsection{Non-intrusive strategies}\label{non-intrusive}
Tools in this category aim to promote effective collaboration by increasing awareness and propagating information (as opposed to actual code) between team members. The guiding principle of these tools is to \textit{react} to potentially emerging conflicts by making team members aware of what's being changed by others, without actually synchronizing their code with the rest of the team. Promoting awareness of changes taking place in the code can lead to an early conflict detection and better communication, ultimately reducing the cost of the resolution. Members of this category include, among others, the following projects.

\subsubsection{Syde \cite{Syde} } establishes team awareness by sharing change and conflict information across software developer’s workspaces by adopting a change-centric approach. Syde provides the notion of synchronous development, where everyone is aware of the activity of others in real time. In order to do so, it extends SpyWare's \cite{robbes2008spyware} change-based software evolution model. The conflicts are classified into two categories: yellow, when there are structural differences between two versions of a node, but none of these versions were checked in the SCM system; red, when there are structural differences between two versions of a node, and one of them
was checked in the SCM system.

\subsubsection{WeCode \cite{guimaraes2012improving}} continuously merges all uncommitted and committed changes inside a software team to create a background system that is analyzed, compiled, and tested to accurately detect conflicts on behalf of developers before check-in. This introduces the case for continuous merging inside the IDE, similarly to the current experience of continuous compilation. WeCode abstractly models the system under construction and the merged system as a tree of typed and attributed nodes, which allows to compare different tree states in order to track changes. Each conflict is reported to the team members that changed the node or nodes affected by it. WeCode deals with four conflict types: structural (e.g., inconsistent node or attribute types), language (i.e., compilation errors), behavior (e.g., potentially unwanted behavior due to unexpected interaction between concurrent changes), and test conflicts (i.e., test methods that fail in the merged system, and its execution flow has two or more methods changed by different members). For structural conflicts, the affected node is that at the location of the conflict in the tree model representing the system's code. For language conflicts, the affected nodes are the ones involved in the compilation error. For behavior conflicts, the affected nodes are those that match the corresponding conflict pattern \cite{guimaraes2012improving}. For test conflicts the affected nodes are the nodes which represent the methods in the failed execution flow. Only the team members that changed the affected nodes and their attributes are notified of the conflict. Such members are found by looking for who changed the affected nodes in the node change tracking information.

\subsubsection{Crystal \cite{brun2011crystal}} provides earlier information that enables more effective behaviors without significant interruption, and in some cases, may help prevent  conflicts from occurring. Crystal can also proactively discover indirect conflicts,  and explicitly checks for compilation and testing conflicts. Crystal unobtrusively reports four kinds of information: the developer’s local state, relationships with other developers or repositories, the possible actions, which is derived from the local state and relationship with the master repository, and guidance about those actions.

\subsubsection{CollabVS \cite{CollabVS}} detects potential conflicts when a software developer starts editing a program element that has a dependency on another program element that has been edited but not checked-in by another developer. It looks for dependencies among three kinds of program elements: file, type (class or interface), and method. Each of these program elements depends on itself. In addition, a type depends on a subtype and supertype, and a method depends on a method it calls or is called by. Such dependencies extend recursively beyond one level. For example, a subtype may have another subtype of its own, and CollabVS works at any depth of such dependencies. Upon detecting a conflict, CollabVS displays a notification balloon that gradually fades away so that in case of a false positive, programmers can ignore it much in the way they ignore junk-mail notifications today. Clicking on the notification balloon automatically takes the user to the conflict inbox displaying a persistent collection of detailed conflict messages regarding the current project. According to CollabVS, the person whose edit created the conflict is responsible for initiating the (possibly collaborative) resolution of the conflict.

\subsubsection{FASTDash \cite{FASTDash}} an interactive visualization that seeks to improve team activity awareness using a spatial representation of the shared code base that highlights team members’ current activities. With FASTDash, a developer can quickly determine which team members have source files checked out, which files are being viewed, and what methods and classes are currently being changed. The visualization can be annotated, allowing programmers to supplement activity information with additional status details. It provides immediate awareness of potential
conflict situations, such as two programmers editing the same source file.

\subsubsection{Lighthouse \cite{Lighthouse}} an Eclipse plug-in that takes the conflict avoidance approach to coordinate developers. Lighthouse distinguishes itself by utilizing a concept called emerging design, an up to date design representation of the
code, to alert developers of potentially conflicting implementation changes as they occur, indicating where the changes have been
made and by whom. The Emerging Design diagram is built dynamically as the developers implement each part of the code, without the need to save or check in the changes made. The diagram is automatically updated with each code change, enabling the developers to always
have an accurate representation of the design as it is currently exists in the developers’ workspaces. The view is updated not only in this developer workspace, but in all developers’ workspaces. Hence, all the developers have the same view of the current design, even if they have not yet checked in or checked out the latest changes.

\subsubsection{Palant\'ir \cite{sarma2003palantir}} a configuration management workspace awareness tool that inverts information flow from pull to push. Instead of informing developers of other efforts only when they themselves perform some configuration management operation (e.g., check in or check out), Palant\'ir increases awareness by monitoring ongoing changes taking place in personal workspaces and continuously sharing information about those changes with developers to whom it is relevant. Rather than excessively notifying developers of conflicts, Palant\'ir inserts small awareness cues in selected parts of the standard Eclipse user interface. The idea is that the cues are unobtrusive, but clearly noticeable at relevant times when, for instance, developers switch from artifact to artifact.

\subsection{Intrusive strategies}
Tools in this category propagate \textit{code} from one team member to another, and their guiding principle is to actively synchronize team members' private code with one another, creating a merged, unified, view of the code. Keeping each individual team member up to date with the unified version assists in \textit{proactively} preventing conflicts (i.e., before they occur), since everyone sees and works on the same (and almost the same), latest, version. Members of this category include, among others, the following projects.

\subsubsection{Collabode \cite{goldman2011collabode}} a web-based Java integrated development environment built to support close, synchronous collaboration between programmers. Programmers use a standard web browser to connect to a Collabode server that hosts their project. The user interface is implemented in HTML and JavaScript and runs entirely within the browser. New programmers can join a project and immediately start working simply by visiting a URL; there is no need to check out code or set up a local development environment. On the server side, Collabode uses Eclipse to manage projects and power standard IDE services: continuous compilation, compiler errors and warnings, code formatting and refactoring, and execution. Any existing Eclipse Java project can be compiled and modified using the Collabode editor (including Collabode itself), with an interface familiar to anyone who has used Eclipse. Collabode addresses the issue of propagating erroneous code states by first giving each programmer a separate, persistent working copy of the program, and then maintaining two versions that integrate programmers' changes from their working copies: a disk version and a union version. The union version is the text that users see and manipulate, and it contains all edits applied by all users, with highlighting and icons to indicate provenance. As long as their methods contain compilation errors, the working copies of team members will each contain only their own method and the disk will contain neither. Once their methods compile, their edits will be shared both with
their collaborator’s working copy, and with the disk version, which corresponds to the content on disk. It is this disk version that is run when either programmer elects to run the program. This version is always free of compilation errors.

\subsubsection{CloudStudio \cite{CloudStudio}} enables developers to work on a shared project repository. Configuration management becomes unobtrusive; it replaces the explicit update-modify-commit cycle by interactive editing and real-time conflict tracking and management. CloudStudio brings flexibility to several new facets of software development, most importantly configuration management (CM): to replace the traditional and painful update-modify-commit reconcile cycle, CloudStudio tracks changes at every location in real time and displays only the selected users’ changes in the integrated editor. The compiler and other tools are aware of the current user preferences, and target the version of the code coinciding with the current view. CloudStudio also integrates communication tools (a chat box and Skype), and includes a fully automated verification component, including both static (proof) and dynamic (testing) tools.

\subsubsection{CSI \cite{2015arXiv150406742L}} an Eclipse plug-in that was our first implementation of the SSD approach \cite{2015arXiv150406742L}, where concurrent changes are forcibly turned into sequential ones, by allowing only one developer to edit any given semantic element (e.g. method) at any given time. Other developers are blocked from concurrently editing that particular element. While blocked they may, however, edit other elements in the code. SSD strives to proactively prevent conflicts by means of fine grained restrictions on element editing. Upon an error free state (i.e., a state where no compilation errors were present), code changes propagate to all team members so as to keep them working on the same, unified, code version. We believe it is highly undesirable for developers to make design related decisions based on stale code, and operated under the fundamental assumption that while coding, a developer would rather wait, obviously, within reason, than engage in a manual merge process incurred by possible code conflicts. CSI's efforts are proactive, directed at preventing conflicts before they actually occur.

\subsection{Intrusive vs. Non-intrusive}
Having worked on implementing CSI \cite{2015arXiv150406742L}, which implemented the general concept of SSD, we have learned that synchronizing code by propagating semantic elements is a very non trivial task, reaching to the darkest and untested corners of the Eclipse IDE \cite{eclipse_ide}, \cite{eclipse_bug}. The fragility of the emerging system, which stemmed mostly from the fact that even the slightest error in the propagation flow could irrevocably bring some, or even all, team members out of sync, brought us back to the drawing table to reevaluate our design prior to moving forward and introducing further complexity. While this problem could be greatly alleviated \cite{slevin_thesis} in the realm of structured editors \cite{quint1994making}, \cite{richy1992multilingual}, they are rarely practiced in today's software development industry, and are mostly restricted to research projects, e.g. \cite{osenkov2007designing}.

While we still believe that the SSD approach has great potential and can be beneficial in more than a few use cases, including the alleviation of merge conflicts, or even the prevention of thereof, we have decided to take some key concepts of CSI, and use them to build a somewhat different, but more robust system. The robustness would come at the cost of shifting from a pro-active approach that attempts to prevent conflicts, to a re-active approach that attempts to boost awareness and promote early conflict detection (see also section \ref{reactiveAndProactive}).

\subsection{Maintaining Consistency}
Some intrusive systems employ one of the following locking schemes to maintain consistency, be it on a textual or a semantic level.
\begin{itemize}
    \item \textbf{Pessimistic locking} takes the view that users are highly likely to corrupt each other's data, and that the only safe option is to serialize data access, so at most one user has control of any piece of data at one time. This ensures data integrity, but can severely reduce the number of concurrent activity the system can support \cite{lockingSchemes}.
    \item \textbf{Optimistic locking} takes the view that such data collisions will occur rarely, so it is more important to allow concurrent access  than to lock out concurrent updates. The catch is that we can't allow users to corrupt each other's data, so we have a problem if concurrent updates are attempted. We must be able to detect competing updates, and cause some updates to fail to preserve data integrity \cite{lockingSchemes}.
\end{itemize}

CloudStudio \cite{CloudStudio} employs a versioning mechanism that resembles to the optimistic locking scheme on a textual level. Since a record is added for every user who edits a line, CloudStudio can search for conflicts by looking up records that only differ in the record's line and author fields. Whenever a user U performs an explicit commit, the base version of the project is updated to reflect U’s latest edits. In case a commit gives rise to a conflict, the base version of that line does not change, otherwise every commit generates a new base version.

CSI \cite{2015arXiv150406742L} implemented a prototype of the SSD \cite{2015arXiv150406742L} methodology, that called for a pessimistic locking scheme on a semantic level \cite{2015arXiv150406741L}, according to which team members were blocked from concurrently editing the same semantic elements (e.g., methods). When a team member tried to edit a method that was already being edited by another team member, he would get an immediate notification in the form of a modal dialog \cite{modalWindowWiki} informing him that the method is locked for editing.

\subsection{Common architectural patterns}
The implementations of systems dealing with code collaboration in general, and merge conflicts in particular, come in several shapes and forms.
\begin{itemize}
    \item \textbf{IDE plug-ins} - software that integrates into an existing IDE, hooking into provided extension points. E.g., Syde \cite{Syde}, Lighthouse \cite{Lighthouse}, Crystal \cite{brun2011crystal}, and WeCode \cite{guimaraes2012improving}.
    \item \textbf{Standalone tools} - software that is meant to be used side by side with the IDE and complement it, rather than integrate with it. E.g., Palant\'ir \cite{sarma2003palantir}, FASTDash \cite{FASTDash}.
    \item \textbf{Customized IDEs} - software that customizes and / or extends an existing IDE's code directly, as opposed to integrate with it via extension points. E.g., CollabVS \cite{CollabVS}.
    \item \textbf{Web-based IDEs} - software that provides a web based UI for interactive code editing. E.g., Collabode \cite{goldman2011collabode}, CloudStudio \cite{CloudStudio}, Cloud9 \cite{cloud9}.
\end{itemize}

\begin{figure}[!htb]
\centering
\minipage[c]{0.45\textwidth}
  \includegraphics[width=\linewidth]{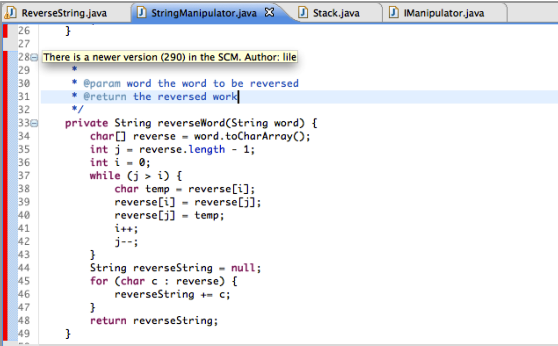}
  \caption{Syde's visual indication for a conflict in a method}\label{fig:syde_conflict_artifact}
\endminipage\hfill
\minipage[c]{0.45\textwidth}%
  \includegraphics[width=\linewidth]{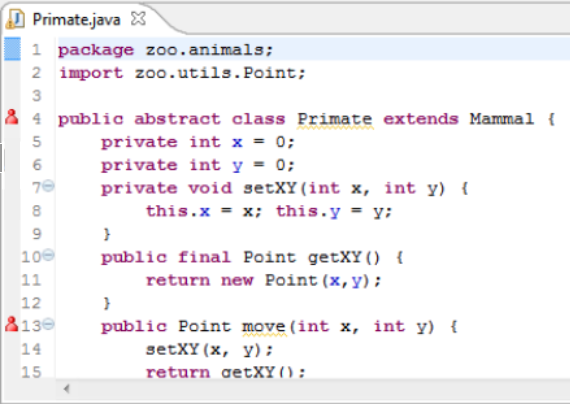}
  \caption{WeCode's visual indication for a conflict in a method}\label{fig:wecode_conflict_artifact}
\endminipage\hfill
\end{figure}

\section{Fine grained visual indications}
Systems that employ non-intrusive strategies tend to integrate with the IDE (i.e., IDE plug-ins) and display visual indications of potential conflicts on the \textit{vertical ruler} (a.k.a. marker bar) of the IDE editor window, manifested as graphical artifacts (see Fig. \ref{fig:syde_conflict_artifact}, Fig. \ref{fig:wecode_conflict_artifact}). While in general such indications make it easier to discover possible conflicts, they do not go into resolutions greater than methods, i.e., they do not decorate specific elements within the scope of a given method. This results in the visual indication being the same whether the potential conflict is in a method's body, parameters, name, and so on. Once a conflict is detected, and a visual indication is displayed to the software developer, it requires further investigation in order to better understand and determine the precise nature of the conflict.

We argue that more fine-grained visual indications can be provided in the editor, making it clearer where the conflict actually lies. In particular, we strive to distinguish between the various fined-grained semantic elements a conflict can be reported on, and to be able to display a visual indication that conveys the essence of that particular conflict in a clearer fashion, and in near real-time.
Following is a list of the fined-grained conflicts applicable in the scope of a method or a field.

\begin{itemize}
    \item Method 
        \begin{itemize} 
            \item Accessibility modifier
            \item Return type
            \item Name
            \item Parameter(s)
                \begin{itemize} 
                    \item Name
                    \item Type
                \end{itemize}
            \item Body
        \end{itemize}
    \item Field
        \begin{itemize} 
            \item Accessibility modifier
            \item Name
            \item Type
            \item Value
        \end{itemize}
\end{itemize}

\begin{figure*}
\centering
  \includegraphics[width=\linewidth]{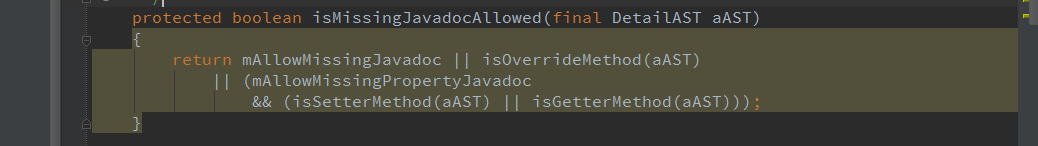}
  \caption{Highlighting a method's body}\label{fig:prototype1}
\end{figure*}

\begin{figure*}
\centering
  \includegraphics[width=\linewidth]{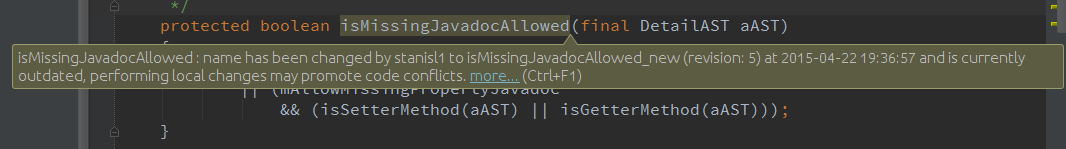}
  \caption{Highlighting a method's name}\label{fig:prototype2}
\end{figure*}

\begin{figure*}
\centering
  \includegraphics[width=\linewidth]{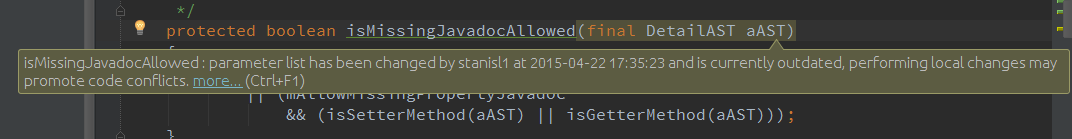}
  \caption{Highlighting a method's parameter}\label{fig:prototype3}
\end{figure*}

\begin{figure*}
\centering
  \includegraphics[width=\linewidth]{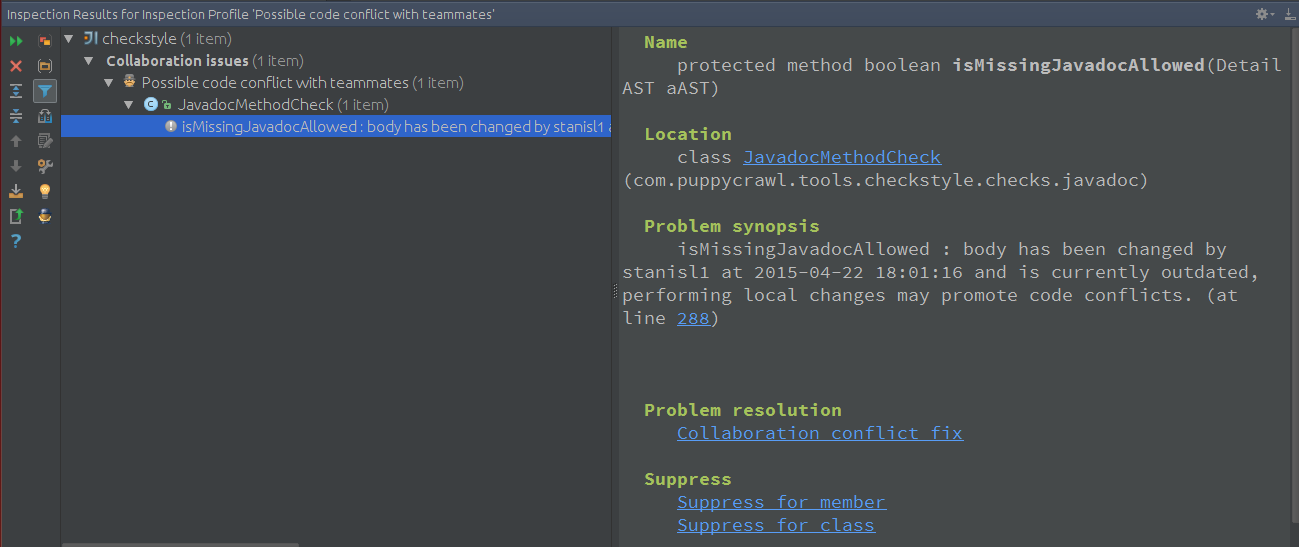}
  \caption{The collaboration inspection in action}\label{fig:prototype4}
\end{figure*}

\section{Change Representation and Conflict Detection}
SpyWare \cite{robbes2008spyware} and Syde \cite{Syde} represent changes as operations on a program's AST. To detect emerging conflicts, every time a new atomic operation is applied to the AST of a developer, it is compared to the ASTs of the others \cite{Syde}. WeCode \cite{guimaraes2012improving} models the system under construction and the merged system as a tree of typed and attributed nodes, where every folder, file, and program element inside a file is a node having a type and a set of attributes (none for folders) in the system’s tree. This allows WeCode to compare two states of a file to determine differences at node and attribute levels. WeCode also detects Language Conflicts \cite{guimaraes2012improving}, i.e., changes that violate the static semantics of the programming language. This is done by automatically compiling the merged version of the system every time it is updated.

While representing changes as AST nodes, or similar tree structures is in many ways intuitive and tempting, it may be somewhat costly to compute and perform diff operations. Compiling the entire codebase upon every change can be a very time consuming task \cite{fowler2006continuous} ("Keep the Build Fast"), especially for large projects. These considerations may challenge the real-timeness of conflict detection mechanisms that rely on these particular techniques.

Our goals were to experiment with more lightweight constructs to represent semantic changes (see section \ref{semanticChanges}) in a program, and gain a better understanding of the trade offs involved in such a decision. We experimented with modeling changes as meta-data entries data holding information pertaining to the specific change made to the code. While all changes have a \textit{semantic path} attribute, some changes may have attributes other do not. A rename change, for instance, holds the old name, as well as the new name attribute, which is absent in a method body change entry. A semantic path has a \textit{semantic path id}, which is a string uniquely identifying a semantic element in the scope of a project, going as high as a method's resolution. For example, if our project's name is Zoo, and we have a class named Zebra residing in a file named Zebra.java with a field name stripeCount, the semantic path id for the stripeCount field is 'Zoo/Zebra.java/Zebra/stripeCount'. A method parameter's semantic path looks like '/project/fileName/className/methodName/paramName/'. The semantic path is a flexible construct, that allows incorporating custom information about semantic elements into the semantic change meta-data.

The task of detecting potential conflicts consists of looking up local and remote semantic changes having the same \textit{semantic path id}, which implies multiple parties have performed changes to be same semantic element. This operation is computationally cheap (O($N \cdot M$), where N is the number of team members, and M is the average number of changes per team member) and greatly speeds up conflict detection, making it near real-time.

\section{The Prototype}
To test our approach we built a prototype, implemented as a plug-in for the IDEA IntelliJ IDE \cite{intellij_homepage}. Its main concerns were as follows.

\subsubsection{Recording local semantic changes}\label{semanticChanges}
This layer is concerned with interpreting editor input as semantic changes. This is achieved by listening to the editor input, and translating the changes made by the software developer into semantic changes. That is, typing the string "int a;" inside a method's body is translated into a \textit{MethodBodyChanged} event. Changing the a field's name from 'aField' to 'theField' would yield a \textit{FieldNameChanged} event.

\subsubsection{Sending and receiving semantic changes}
States where the code fails to be successfully compiled, may yield incorrect semantic interpretations due to parsing errors. This layer is responsible for detecting error free states and utilizng them in order to filter, consolidate, and propagate the relevant semantic changes recorded so far, to the collaborating parties. While periodically compiling the code might be the most straight forward option to detect error free states, it quickly became apparent it was not feasible even for small projects. A more performant solution to this problem was harnessing the power of the \textit{on-the-fly} error detection of the IDE itself \cite{intellij-onthefly}, \cite{intellij_inspections}. In addition, this layer also integrates with the SCM in order to compare the version of the incoming semantic changes, with that of the local semantic changes, and rejects the former in case it is lower. This was a design choice made in order to prevent semantic changes originating from outdated parties from showing up as conflicts at the up-to-date parties (we deemed it reasonable, since in such a case it should be the outdated parties' responsibility to resolve any conflicts. This is also supported by the fact the outdated parties will be obligated by the SCM system to "update" their code before they can perform a check-in. Receiving changes with higher versions is allowed, and may help outdated parties to be made aware of conflicts stemming from the changes that they made on stale code, after other team members had continued working on newer code that had a higher version number, and was already checked-in to SCM repository, effectively making it the source of truth.

\subsubsection{Detecting potential conflicts and displaying indications}\label{detecting_conflicts}
\begin{figure*}
  \includegraphics[width=\linewidth]{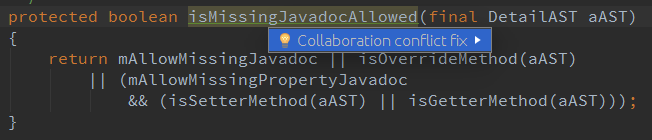}
  \caption{The collaboration inspection's quick fix popup}\label{fig:quick_fix_popup}
\end{figure*}

\begin{figure*}
  \includegraphics[width=\linewidth]{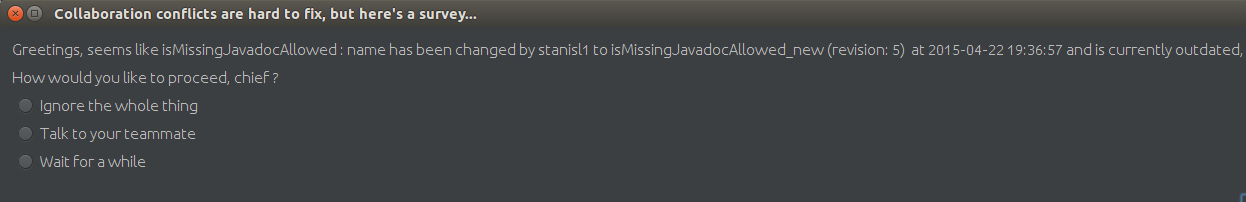}
  \caption{An IDE-native survey embedded in the collaboration inspection's quick fix flow}\label{fig:quick_fix_survey}
\end{figure*}

Once the plug-in has obtained both local, and the collaborating parties' semantic changes, it can compute the merge conflicts by comparing the semantic change events. To achieve the goal of displaying a fine-grained visual indications we leverage the inspection mechanism of IDEA IntelliJ \cite{intellij_inspections}. By implementing a new kind of code inspection, we were able to hook into the IntelliJ framework, and gain access to the PSI (Program Structure Interface) \cite{intellij_psi} where we could specify the exact element that has undergone conflicting changes, and thus have it decorated with a proper graphical artifact.
We display two kinds of visual indications. The first (yellow background highlighting) is displayed when one of the parties makes a change to a certain element, while other parties have not changed that element yet. This indication is aimed to serve as a soft warning, enhancing the awareness of what others are changing. 
The second (squiggles) is displayed when multiple parties have made concurrent changes to the same element. This indication implies a potential merge conflict might be present. 
Both code inspections suggest a "Quick Fix" (see Fig. \ref{fig:quick_fix_popup}), where an IDE-native survey dialog is displayed (see Fig. \ref{fig:quick_fix_survey}), asking the software developer to specify his intended action in light of the possible conflict. It does not actually fix the conflict, and was meant to allow for subjects to provide feedback for research purposes in a native manner, and as part of their coding flow.

\subsection{Limitations}
\label{sec:limitations}

In its current design, certain cases may produce false positives due to the fact we only store some of the information pertaining to a given change. For example, currently we do not track changes made inside a method's body, which limits our ability to reason about statement level conflicts in this scope. Our support for undo operations is also limited at this stage, while we are able to monitor SCM revert operations performed on a file, and react by deleting the relevant local semantic changes, text undo operations do not result in removing the semantic changes they originally triggered. In some cases this may lead to reporting false conflicts. 
The current implementation currently supports teams of two, but was designed with supporting the general case of N members in mind.
It is important to stress that these limitations are not inherent in the suggested approach, but rather in its current implementation state.

\section{Evaluation}

\subsection{Experiment Design}

To evaluate our approach and prototype we designed an experiment carried out in pairs, co-located in a single room. Subjects could communicate at all times, without any restrictions. Each team member received a list of three tasks (see appendix \ref{app:tasks}), where each task asked to perform certain changes to the CheckStyle \cite{checkstyle} open source project. The tasks were arranged in a way that would require team members to work on the same method(s) at least some of the time. While working on their tasks, subjects were presented an IDE-native survey (see section \ref{detecting_conflicts}) whenever they attempted to fix a potential conflict.
Prior to working on these tasks, subjects were given a brief background about the prototype plug-in and a short demonstration. During the coding session, subjects' screens were video-captured (using \cite{SimpleScreenRecorder}), and their IDE logs recorded.
After the coding session, an interview inspired by \cite{blackwell2000cognitive} was conducted with each subject (see appendix \ref{app:interview}), after which they were also given an opportunity to provide their own free-style feedback.

\subsection{Preliminary Results}

We have conducted 3 evaluation sessions, with highly experienced software developers, whose experience as professionals in the software industry (in years) distributed as follows, with respect to our pairing: (2,9), (6,12), and (11,11). All the subjects had previous experience with IntelliJ.

We feel that the integration with IntelliJ, and employing well known mechanisms such as code inspections made it easier to both develop the prototype plug-in, and deliver information to the software developers using it. The familiar look and feel of squiggles and background highlighting (see Fig. \ref{fig:prototype1}, Fig. \ref{fig:prototype2}, Fig. \ref{fig:prototype3}) were something software developers were no strangers to.

Subjects reported that the presence of visual indications pertaining to potential conflicts (a.k.a, collaboration inspections, see Fig. \ref{fig:prototype4}), made them think of what their team member was doing and prompted them to communicate in the face of indications of a potential conflict. This was also supported by their answer "Talk to your teammate", which was often provided to the question "How would you like to proceed?" that was presented to them if they had clicked the "Collaboration conflict fix" option. Subjects also mentioned that they had expected the visual indication to go away once they closed the "Collaboration conflict fix" dialog, and found it annoying and even disturbing that it did not. In fact, their discontent with the persistent highlighting (i.e., the visual indications) was sometimes expressed by answering the subsequent "How would you like to proceed?" (in light of a potential conflict) questions with "Ignore the whole thing". This can indicate frustration on their behalf and imply that while visual indications have their merits, they should not be repetitive or too visually dominant so as to not over impose themselves on the IDE user (i.e., software developer), who at some point, can break and start ignoring them altogether.

One subject suggested that perhaps the fact visual conflict indications did not go away easily and made it harder to read big chunks of code, could encourage software developers to make their methods shorter. While shorter methods are indeed a higher goal, we feel that our prototype might not be the ideal tool to effectively achieve it.

Few subjects mentioned it would be nice to be able to view the changes made by their team members in real time, in addition to just being indicated that a given element was changed. Some took it even further and said it might be nice to be able to "accept" the changes of a team member, and resolve the conflict on the spot.

When asked if they had ever used a similar system (to the prototype plug-in), most subjects replied they had not, except for one, who said that in his view, a SCM system is kind of the same except for the real-time indications, and he uses SCM systems all the time.

Some subjects also reported that their attitude towards potential conflicts could be somewhat different in a real world scenario, both because potential conflict in production code has much greater implications than in a synthesised experiment, and because they had a preconceived (speculative) idea about what the plug-in "intended" to have them do when a conflict was indicated.

\section{Conclusions and Further Work}

Conflict detection puts forth numerous challenges, of which the prominent are: detecting conflicts with minimal false positives, and doing so in near real-time. 
Existing approaches have suggested several methods to deal with conflict detection, some concentrated on representing changes as AST nodes, or similar tree structures, others chose to continuously merge-compile the code in the background. Both techniques tend to be costly in terms of computation power, (especially when dealing with large a codebase) which can have an adverse effect on the time it takes to detect a potential conflict and act upon it. 

We propose a novel approach, where semantic changes are represented in a more lightweight fashion so as to allow more real-time detection, and can be used to provide software developers with fine-grained, visual information on a semantic level about potential merge conflicts. 
We have implemented a prototype plug-in for the IDEA IntelliJ IDE and conducted a qualitative evaluation, which indicates that the presence of visual indications in the IDE enhances team members' awareness of potential merge conflicts, and their willingness to communicate and discuss it. Our IDE-native survey (see section \ref{detecting_conflicts}) suggests that once prompted, team members are willing to communicate with each other if they believe a potential conflict might be present in the system. Using fined-grained indications, this communication can be made very effective.

The limitations that have been described (see section \ref{sec:limitations}) are not inherent to the approach itself, but rather to its current implementation. We believe that future versions could relieve some, if not all, of them. We also feel that this approach has the potential to meet the key requirements of delivering near real-time, highly precise conflict diagnostics and convey it clearly by pointing out specific fined-grained semantic elements, a feature that is currently missing in similar systems to the best of our knowledge.

It could be beneficial to explore additional collaboration features, such as be able to inspect other team members' changes and compare them to the local ones, or even accept / reject remote changes. The realm of exchanging code changes in an effective, controlled, real-time manner is a frontier yet to be conquered.

Evaluating our approach and prototype in a setting with more than two software developers is also a matter of interest, and may shed further light on the nature of conflicts in a more real-world like scenarios, where teams tend to have more than two members.

The trade off between being able to report conflicts in real-time, at the cost of potentially increasing false positives is an interesting question in its own right, and further research is needed in order to better understand its affect on software developer's productivity.

\section*{Acknowledgements}
This research was supported by THE ISRAEL SCIENCE FOUNDATION, grant No. 476/11.

We would like to thank Professor Shmuel Tyszberowicz for his useful comments on the draft of this paper.

\clearpage
\onecolumn
\appendix
\renewcommand{\thesubsection}{\Alph{subsection}}

\subsection{Coding Tasks}
\label{app:tasks}

All coding tasks were performed on CheckStyle \cite{checkstyle}.

\begin{itemize}
\item Software Developer 1
    \begin{enumerate}
        \item Open JavadocMethodCheck.java.
        \item Replace the first while loop in the “checkComment” method with a functional style block.
        \item Replace the “paramIt” loop in the “checkParamTags” method with a function style block.
        \item Replace the “it” loop in the “checkReturnTag” method with a function style block.
    \end{enumerate}
\end{itemize}

\begin{itemize}
\item Software Developer 2
    \begin{enumerate}
        \item Open JavadocMethodCheck.java.
        \item Replace the second while loop in “checkComment” with a functional style block.
    \item Replace the “typeParamsIt” loop in the “checkParamTags” method with a function style block.
        \item Replace the “tagIt” loop in the “checkThrowsTag” method with a function style block.
    \end{enumerate}
\end{itemize}

\subsection{Interview questions}
\label{app:interview}

\begin{enumerate}
\item Have you used other similar systems? (If so, please name them)
\item What would you say is your level of proficiency in using the system?
\item What actions did the system make easier to perform?
\item What actions did the system make harder to perform?
\item How did the system affect your productivity?
\item How did the system affect your collaboration with your teammate? Why?
\item Did you want to communicate with your teammate during the tasks?
\end{enumerate}

\clearpage
\twocolumn
\bibliographystyle{IEEEtran}
\bibliography{bibliography}

\end{document}